\documentstyle[editedvolume]{crckapb} 

\input epsf
\def\Mpc{{$h^{-1}\,$Mpc}}
\def\kms{{km$\,$s$^{-1}$}}
\begin{opening}
\title{Void Hierarchy in the Northern Local Void}
\subtitle{Faint structures in low density regions of the nearby Universe}
\author{U. Lindner}
\author{K.J. Fricke}
\institute{Universit\"ats--Sternwarte, G\"ottingen, Germany}
\author{J. Einasto}
\author{M. Einasto}
\institute{Tartu Astrophysical Observatory, T\~oravere, Estonia}
\end{opening}
\runningtitle{Void Hierarchy in the NLV}
\begin{document}

\section{Introduction}
Empirical studies of the Large--Scale Structure 
in the nearby Universe come in two complementary modes,
namely the investigation of either the distribution of
luminous matter or voids:
{\bf (i)} The description of the galaxy and cluster distribution
employs correlation functions, clustering analysis, 
topological methods, et cetera.
{\bf (ii)} The investigation of the empty regions between 
systems of galaxies uses void probability functions, 
mean diameters of voids, the compilation of void catalogues,
and so forth.

\begin{figure}\epsfysize=10.5cm
\vskip -1.0cm
\epsffile{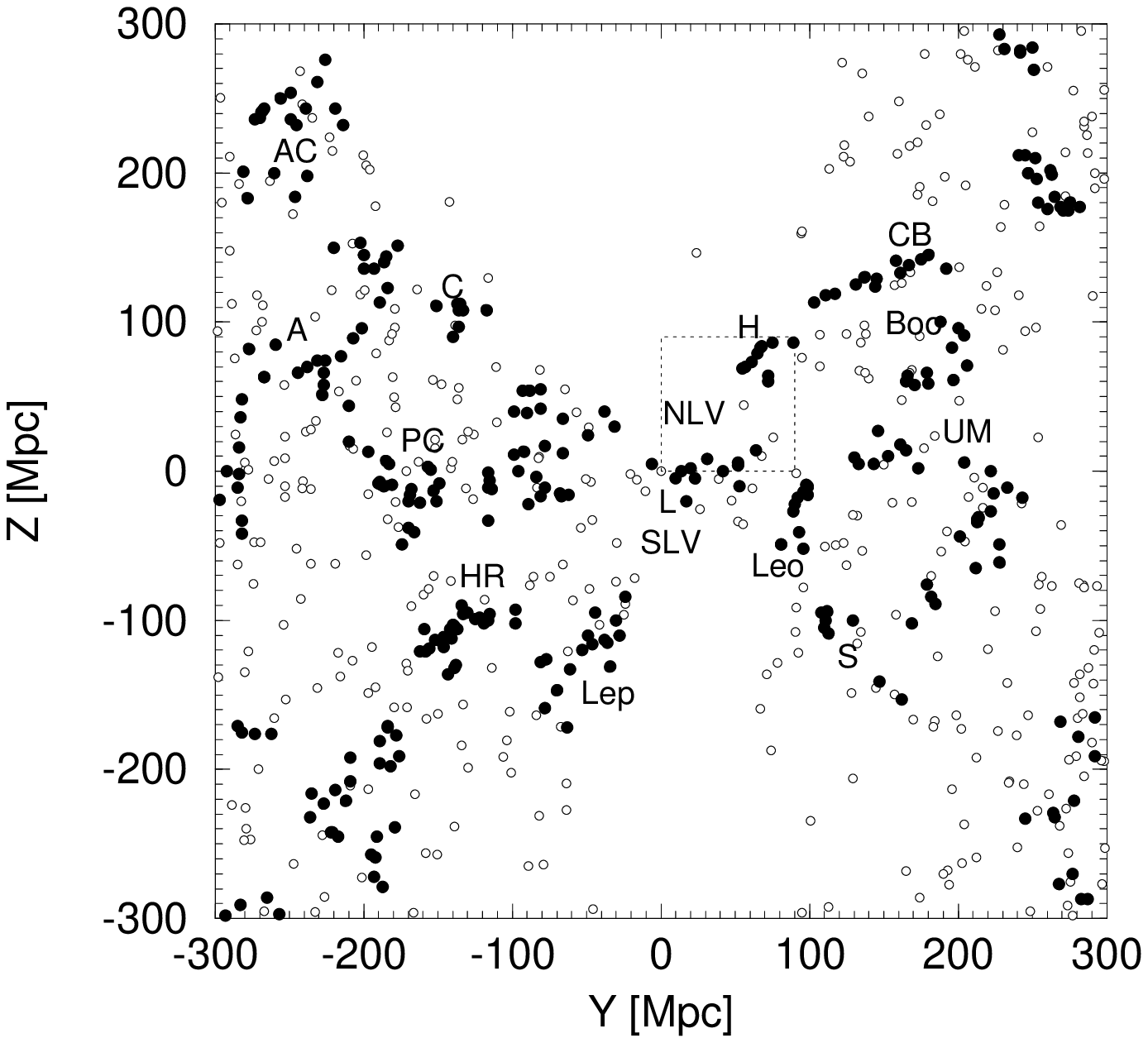}
\vskip -1.5cm
\end{figure}

\begin{figure}\epsfysize=8.0cm
\centerline{\epsffile{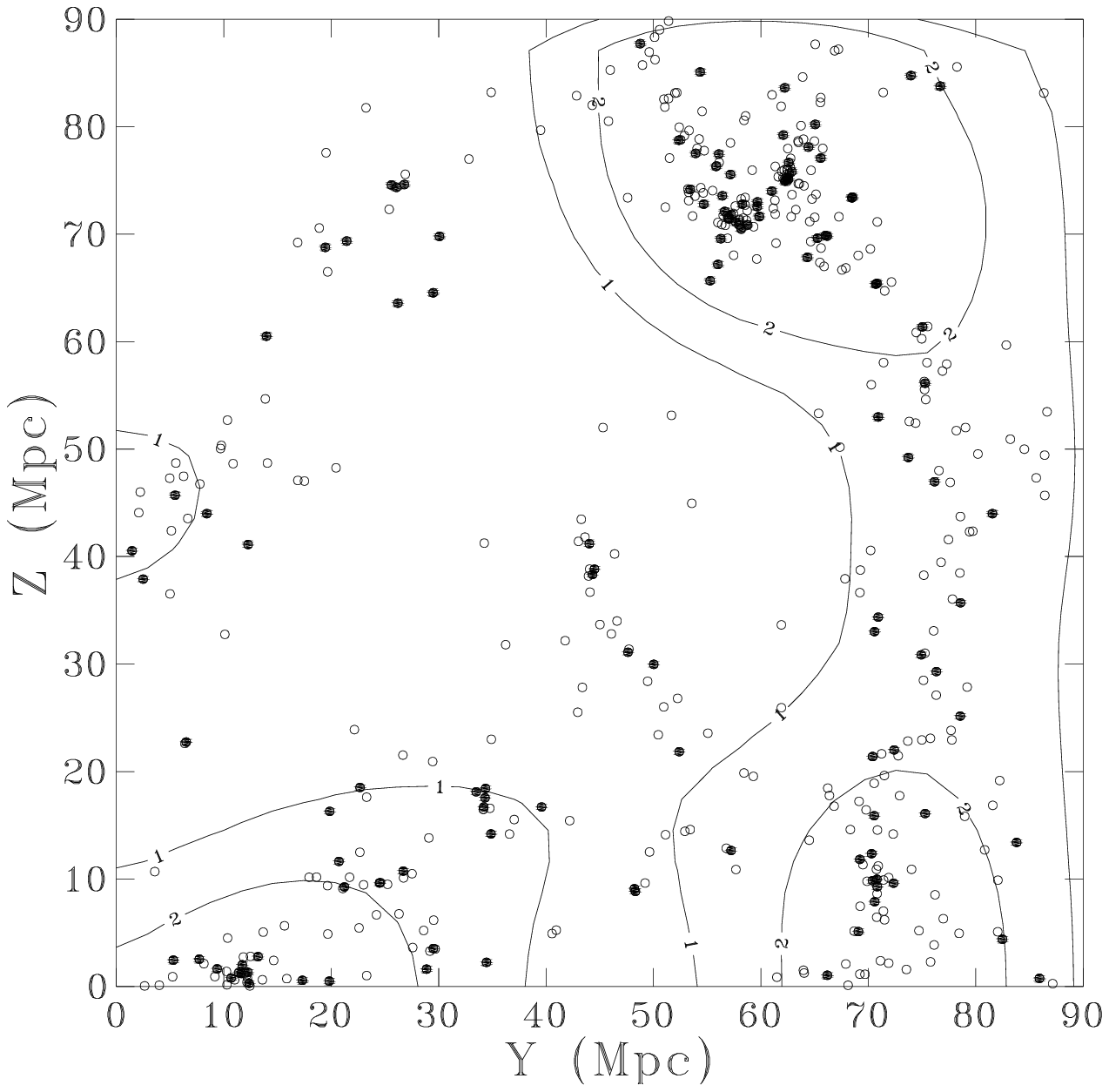}}
\caption{a) rich (Abell--ACO) clusters in supergalactic
coordinates X, Y, Z; sheet is $\Delta\,$X = 200 \Mpc\ thick (M.
Einasto 1995);
\quad b) galaxies in the NLV region (indicated by
the square in Figure~1~a)}
\end{figure}

Here we focus on the study of voids and for this purpose it
must be clarified what voids are.
Two essentially different definitions are in use: 
Voids can be defined as 
{\bf (i)} regions of low density in a suitably smoothed
density field of the galaxy distribution 
or as 
{\bf (ii)} regions completely empty of a certain type of 
object, e.g. rich or poor galaxy clusters or galaxies of a specified
morphological type or luminosity limit.

Figure~1$\,$a) shows the distribution of rich galaxy clusters
from the Abell--ACO catalogue which constitute the cores of
superclusters. The large voids between superclusters are called
{\it supervoids}. They are by definiton devoid of rich clusters
and have diameters of about 100 \Mpc . They are the largest known
voids in the Universe. 
Superclusters are connected and form a chess board like regular
network. 

Figure~1$\,$b) gives a closer look 
to the Northern Local Void (NLV) region indicated in Figure~1$\,$a). 
This closest supervoid which is bordered
by the Hercules--, Coma-- and Local Supercluster is not completely
empty but contains systems of galaxies of various richness.
This finding motivates a more detailed study of the Large-Scale
structures in this low density region concerning morphological 
type and luminosity of galaxies.

\section{Basic Data}
The region of the Northern Local Void (NLV)
covers $12^h <\alpha < 18^h$, $\delta > 0^{\circ}$ on the sky
and is about 15000 \kms\ deep in redshift space.
Using a Hubble constant of H $= h\,100\,$km$\,$s$^{-1}\,$Mpc$^{-1}$
this corresponds to a distance of 150 \Mpc .

Redshifts for poor clusters of galaxies from the Zwicky et al. (1961--68) 
catalogue have been estimated by E. Tago (1993).
Galaxy redshifts are taken from the ZCAT (Huchra, 1994) and various 
other sources, mainly the Arecibo 21 cm survey (Giovanelli et al., 1992).
From this redshift compilation which is complete up to $m_{lim} = 15.5$
we selected three absolute magnitude limited
cubic samples of different size (L = 60, 90 and 120 \Mpc ) for galaxies
of early type (E and S0) and all morphological types. Using
the completeness limit $m_{lim} = 15.5$
the luminosity limits of the samples are $M_{lim}$ = --18.8, --19.7 and
-20.3, respectively.

\begin{figure}\epsfysize=10.75cm\vskip -0.5truecm
\hskip -1.0truecm\epsffile{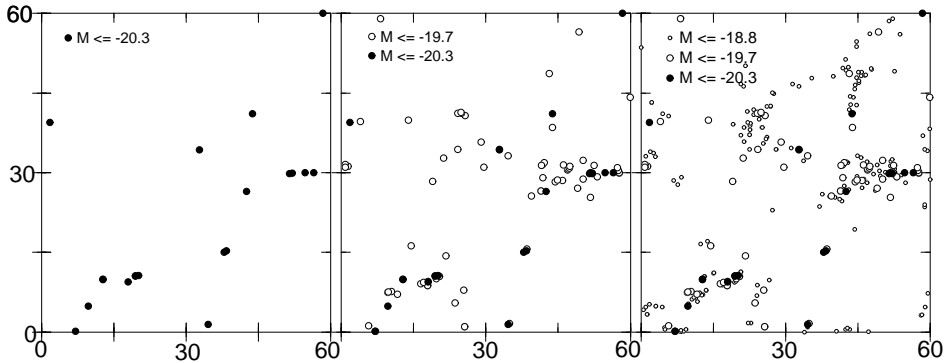}\vskip -5.3truecm
\caption{Cross sections through cubic samples of galaxies with 
different luminosity limit (cf. legend) in the NLV region. The
cube size is $L = 60\,$\Mpc .}\end{figure}

In Figure~2 we present cross sections through cubic samples 
($L = 60\,$\Mpc ) for galaxies of all types with different 
luminosity limit (as indicated in the legend).
Bright galaxies are situated in the centers of galaxy systems.
Fainter galaxies gather around these centers and in some case
they form new subtle structures (filaments or sheets) between these systems.

For detailed quantitative studies we chose a void analysis
applying the empty sphere method which is particularly sensitive to subtle
structures in the galaxy distribution.
For a more detailed description of the methods we refer the reader
to Lindner et al. (1995).

\begin{figure}\epsfysize=10.0cm\vskip -4.0truecm\hskip -0.25truecm
\epsffile{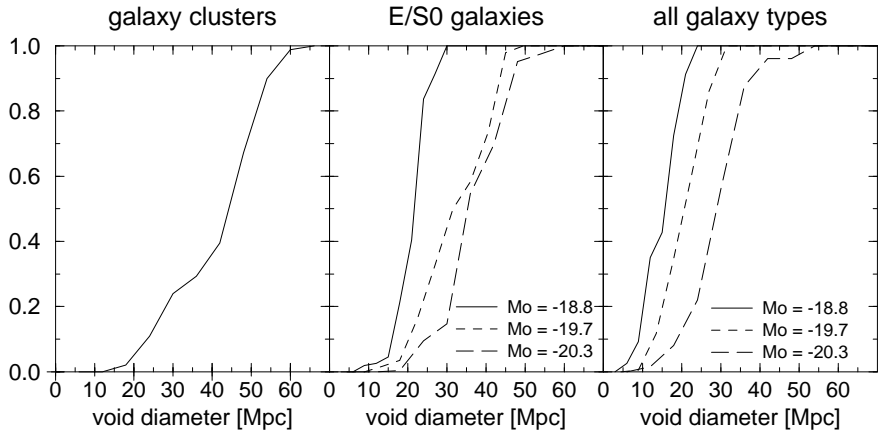}\vskip -0.5truecm
\caption{Cumulative distributions of void diameters}\end{figure}

\section{Statistics of void diameters}
In Figure~3 we present cumulative distributions of void diameters
for Zwicky clusters, early type galaxies 
and galaxies of all morphological type.
Three different luminosity limited galaxy samples are indicated.
We see that voids surrounded by early type galaxies
are  larger than those found in respective samples of all
morphological types and fainter galaxies outline smaller
voids than bright galaxies. The distribution of void
diameters in the case of clusters is similar to that of
bright elliptical galaxies resembling the fact
that these galaxies reside predominantly in cluster cores.
This dependence of void diameters on the type of object and
the galaxy luminosity is quantified by the mean values
given in Table~1 with random errors $\epsilon$ and the rms 
scatter $\sigma$.

\begin{table}[htb]
\begin{center}
\caption{Mean diameters of voids $\langle D\rangle$ 
surrounded by different objects}
\begin{tabular}{rlr}
\hline
type of object \qquad\qquad & $\qquad\langle D\rangle \pm\epsilon$ & $\sigma$\qquad\qquad \\
\hline
 rich clusters (Abell/ACO)           & $\approx$~100~~ \Mpc & \\
 poor clusters (Zwicky)              & 37.0$\pm$3.5 \Mpc & 11.7 \Mpc \\
 elliptical galaxies ($M \le -20.3$) & 36.4$\pm$3.4 \Mpc & 9.2 \Mpc \\
 galaxies brighter than $M_0 = -20.3$  & 25.7$\pm$1.3 \Mpc & 6.4 \Mpc \\
 galaxies brighter than $M_0 = -19.7$  & 18.8$\pm$0.9 \Mpc & 5.3 \Mpc \\
 galaxies brighter than $M_0 = -18.8$  & 14.0$\pm$0.9 \Mpc & 4.0 \Mpc \\
\hline
\end{tabular}
\end{center}
\end{table}

The number of objects in various samples is very different,
ranging from 1850 faint galaxies to 140
bright elliptical galaxies. Hence the question arises whether
our results reflect a real property of the galaxy distribution
or if they are caused only by a different number of objects.
A comparison study to Poisson samples reveals statistic
significance of our results (cf. Lindner et al. 1995).

\section{Void catalogues}
Void catalogues have been compiled
by various authors: Kauffmann \& Fairall (1991) found cubic
empty regions; Slezak, de Lapparent \& Bijaoui (1993) used
a wavelet method to detect low density regions and
El--Ad, Piran \& da Costa (1996) presented a new void search
algorithm for automated void detection classifying galaxies
into wall and field galaxies, the latter being allowed
to lie inside voids.
But all these investigators did not consider
the different luminosity of galaxies.

We employed the empty sphere method to compile void 
catalogues for three different absolute magnitude limited galaxy samples. 
They are named A, B and C for $M_{lim} = -20.3, -19.7$ and $-18.8$,
respectively, and contain the position and distance of void centers
and the void diameter.

About one third of the voids in catalogue A can be identified 
also in catalogue B and a similar result is found for
catalogues B and C. The study of overlappings between
voids from different catalogues shows that
{\bf (i)} systems of faint galaxies enter into voids outlined
by bright galaxies and make their diameter smaller or that 
{\bf (ii)} one or more faint galaxy systems split large voids into several
subvoids.

\section{Void hierarchy}
Our statistical studies reveal that
void diameters depend on the type of object surrounding voids.
Cluster defined voids are the largest voids and voids surrounded by
early type galaxies are larger than those formed by galaxies
of all morphological type. Furthermore we found that voids
are smaller in samples of fainter galaxies.

Intersections between voids from different
catalogues show that voids outlined by bright galaxies 
contain fainter galaxies and may be interlaced by systems
of fainter galaxies which divide them into smaller subvoids.

These results indicate that voids form a hierarchical system 
resembling the hierarchy of galaxies, clusters of galaxies and
superclusters.

\begin{table}[htb]
\begin{center}
\caption{The hierarchy of voids}
\begin{tabular}{l}
\hline
voids surrounded by rich clusters: supervoids \\
voids surrounded by poor (Zwicky) clusters of galaxies \\
voids surrounded by bright elliptical galaxies \\
voids surrounded by galaxies brighter than $M_0 = -20.3$ \\
voids surrounded by galaxies brighter than $M_0 = -19.7$ \\
voids surrounded by galaxies brighter than $M_0 = -18.8$ \\
voids surrounded by galaxies fainter than $M_0 = -18.8$ (dwarf galaxies) ? \\
\hline
\end{tabular}
\end{center}
\end{table}

As far as we know the upper limit of this hierarchy is
given by the supercluster network with
supervoids of about 100 \Mpc\ in diameter.
By now the lower limit is not well determined. In our investigation the
smallest voids are defined by galaxies from complete samples
with absolute magnitude limit $M_{lim} = -18.8$.
Studies of incomplete samples fainter than this limit give
strong hints that the void hierarchy may continue to
dwarf galaxies. That means dwarf galaxies are not distributed
homogeneously but follow the structures delineated by
bright galaxies or form subtle filamentary 
substructures between dense galaxy systems (cf. Lindner et al. 1996).

\section{Conclusion}
Numerical simulations of voids by
Reg\"os \& Geller (1991), van de Weygaert \& van Kampen (1993),
Dubinski et al. (1993), Sahni et al. (1994)
agree with our picture of the present internal structure of 
supervoids found in the galaxy distribution of the NLV.
They studied the evolution of structures inside voids
and found that small (primordial) voids merge in course of time to form
larger voids inside supervoids at later epochs, i.e. there was more
substructure in the beginning  than today.
The existence of hierarchical substructures 
in supervoids rules out the non hierarchical hot dark matter model.
The existence of structure at small and 
large scales favors a double power law power spectrum for 
primordial density fluctuations, as suggested by Einasto et al.
or some mixed dark matter model.

Other, more distant supervoids may have
similar internal structure. For example Szomoru et al. (1993, 1996) 
have found some galaxies in the Bo\"otes Void 
which may be part of a similar web as we have found in the NLV.
In general selection effects hinder to detect the full
extend of faint structures.
Deeper investigations like the 
ESO slice project (Vettolani et al. 1996) and
Las Campanas Redshift Survey (Shectman et al. 1996) 
confirm the existence of a supercluster -- supervoid network
and faint structures inside the supervoids.
The next generation redshift surveys
may help to reduce the problems connected with selection effects
and to establish the existence and lower limit of the 
void hierarchy.

\end{document}